\documentclass[prl,superscriptaddress,twocolumn]{revtex4-2}
\usepackage[utf8]{inputenc}
\usepackage[T1]{fontenc}

\usepackage{amsmath}
\usepackage{amssymb}
\usepackage{newtxtext}
\usepackage[smallerops]{newtxmath}
\let\lambda\lambdaup
\allowdisplaybreaks
\usepackage{graphicx}
\graphicspath{{fig/}}
\usepackage{xcolor}
\usepackage[colorlinks=true,linkcolor=blue,citecolor=blue,urlcolor=blue]{hyperref}
\usepackage[capitalise]{cleveref}

\begin{document}

\title{Fragmented spin ice and multi-$k$ ordering in rare-earth antiperovskites}

\author{Attila Szab\'{o}}
\address{Rudolf Peierls Centre for Theoretical Physics, University of Oxford, Oxford OX1 3PU, UK}
\address{ISIS Facility, Rutherford Appleton Laboratory, Harwell Campus, Didcot OX11 0QX, UK}
\author{Fabio Orlandi}
\address{ISIS Facility, Rutherford Appleton Laboratory, Harwell Campus, Didcot OX11 0QX, UK}
\author{Pascal Manuel}
\address{ISIS Facility, Rutherford Appleton Laboratory, Harwell Campus, Didcot OX11 0QX, UK}

\date{\today}

\begin{abstract}

We study near-neighbour and dipolar Ising models on a lattice of corner-sharing octahedra. 
In an extended parameter range of both models, frustration between antiferromagnetism and a spin-ice-like three-in-three-out rule stabilises a Coulomb phase with correlated dipolar and quadrupolar spin textures, both yielding distinctive neutron-scattering signatures.
Strong further-neighbour perturbations cause the two components to order independently, resulting in unusual multi-$k$ orders.
We propose experimental realisations of our model in rare-earth antiperovskites.

\end{abstract}

\maketitle

%\section{Introduction}

Spin ice has been a paradigmatic example of frustrated magnetism since its experimental realisation in the rare-earth pyrochlores $\mathrm{Dy_2Ti_2O_7}$ and $\mathrm{Ho_2Ti_2O_7}$~\cite{Harris1997GeometricalHo2Ti2O7,Ramirez1999Zero-pointIce,Bramwell2001SpinMaterials}.
It is defined on a lattice of corner-sharing tetrahedra by a local two-in-two-out (2I2O) constraint on each tetrahedron.
This constraint results in extensive ground-state degeneracy~\cite{Pauling1935TheArrangement,Anderson1956OrderingFerrites,Ramirez1999Zero-pointIce,denHertog2000DipolarMagnets}, long-range dipolar correlations~\cite{Bramwell2001SpinSystem,Isakov2004DipolarMagnets,Henley2005Power-lawAntiferromagnets}, and fractionalised excitations that behave as emergent magnetic monopoles~\cite{Hermele2004PyrochloreMagnet,Castelnovo2008MagneticIce,Morris2009DiracDy2Ti2O7,Fennell2009MagneticHo2Ti2O7,Bramwell2009MeasurementIce,Dusad2019MagneticNoise},
a phenomenology known as the \textit{Coulomb phase}~\cite{Henley2010TheSystems,Castelnovo2012SpinOrder}.
Spin ice, and derivatives such as quantum spin ice~\cite{Hermele2004PyrochloreMagnet,Gingras2014QuantumMagnets,Rau2019FrustratedPyrochlores} and monopole-crystalline fragmented spin ice~\cite{Lefrancois2017FragmentationInjection,Cathelin2020FragmentedDy2Ir2O7,Pearce2022MagneticIridates}, are among the best-studied frustrated magnets
due to the abundance of rare-earth pyrochlore materials with large local moments~\cite{Wiebe2015FrustrationOxides}.

\begin{figure}
    \centering
    \includegraphics{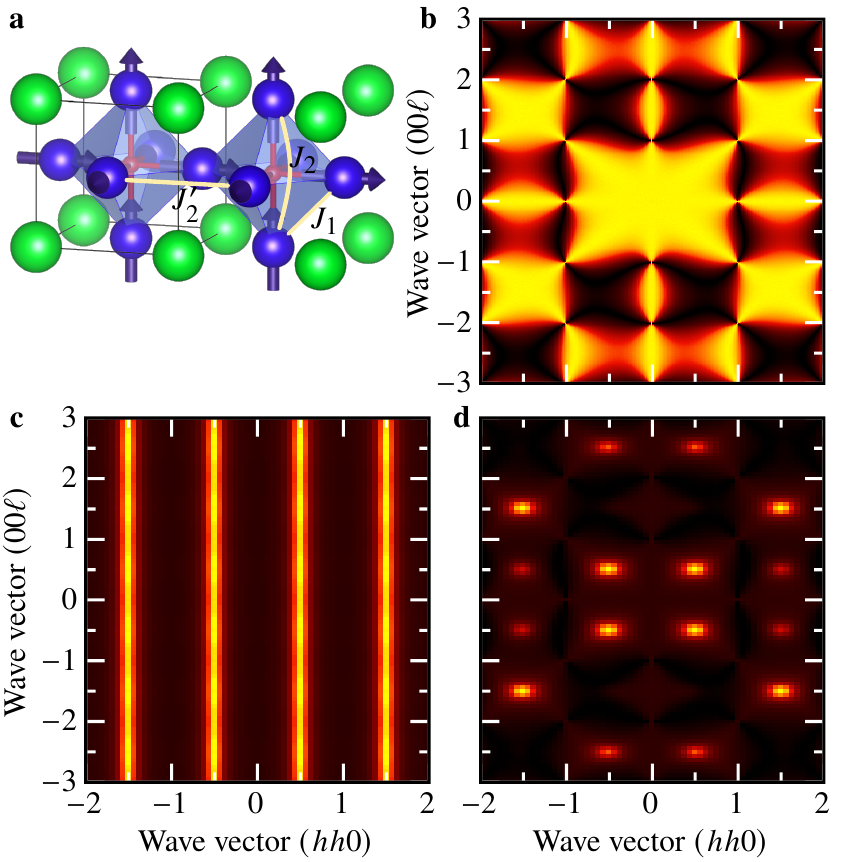}
    \caption{
    (a) Two unit cells of the antiperovskite structure showing the local $\langle100\rangle$ Ising axes and near-neighbour exchange couplings $J_1$, $J_2$, $J_2'$ of the magnetic ions (blue).
    (b) Spin-flip (SF) neutron-scattering pattern of~\eqref{eq: ideal J1-J2 Hamiltonian} at $T\ll J$ for neutrons polarised along $[1\bar10]$. Sharp pinch points are visible at all $\Gamma$ points of the simple cubic lattice. [The non-spin-flip (NSF) structure factor is independent of wave vector.]
    NSF (c) and SF (d) neutron-scattering pattern of~\eqref{eq: general J1-J2 Hamiltonian} for $J_2<J_1$ at $T\to0$.
    In addition to pinch points, rods of diffuse scattering arise along wave vectors of the form $(1/2,1/2,\ell)$, which are fully visible in the NSF pattern, and yield smooth maxima in the SF one.
    }
    \label{fig: ideal ice pinch points}
\end{figure}

Nevertheless, Coulomb phases are not limited to the pyrochlore lattice, as shown by the rich physics of (artificial) square ice~\cite{Nisoli2013Artificial} and kagome ice~\cite{Zhao2020Kagome} in two dimensions. 
A three-dimensional alternative would be a lattice of corner-sharing octahedra (e.g., the edge midpoints of a simple cubic lattice), endued with Ising spins pointing into or out of these octahedra and a local three-in-three-out (3I3O) constraint, defining a twenty-vertex model~\cite{Baxter1982ExactlyMechanics}; 
understanding this rule as a local divergence-free constraint yields a Coulomb phase similar to pyrochlore ice.
This geometry is realised in the cubic phase of antiperovskites in which rare-earth metal ions with strong easy-axis anisotropy form coordination octahedra around a central anion (\cref{fig: ideal ice pinch points}a).
Such models have received some theoretical attention~\cite{Hermele2004PyrochloreMagnet,Benton2021HigherOrder,Sklan2013Nonplanar}
but have not yet been considered as an experimentally viable platform for spin ice, as realising the perfect Coulomb phase requires significant fine-tuning.

Our work addresses this gap by studying both near-neighbour exchange and dipolar Ising Hamiltonians on the antiperovskite lattice. 
For the former, we find an extended parameter range in which frustration between antiferromagnetic alignment and the 3I3O rule brings about a new kind of Coulomb phase.
Octahedra in its ground-state manifold show correlated dipolar and quadrupolar alignment, both of which give rise to distinctive features in the magnetic structure factor.
This phase, dubbed \textit{fragmented spin ice,} is further stabilised by dipolar interactions against weak further-neighbour interactions.
Stronger perturbations result in a variety of ordered phases: in those derived from fragmented ice, the dipolar and quadrupolar components order independently, resulting in unusual multi-$k$ structures.

Our results mark out rare-earth antiperovskites, with large Ising-like moments, as well as structural analogues~\cite{Coates2021,Overy2016StructuralIce}, as a novel platform for realising Coulomb-phase physics and may explain experiments that found no magnetic ordering in such materials down to the lowest temperatures~\cite{Hoppe2002GdAntiperovskite,Kloss2017HoAntiperovskite}.
Moreover, they demonstrate that the mechanisms underlying dipolar pyrochlore spin ice~\cite{Isakov2005WhyRules,Melko2004MonteModel,denHertog2000DipolarMagnets} are generic and may help stabilise more exotic icelike models~\cite{LantagneHurtubise2018ThinFilms,Yan2020Rank2Breathing,Benton2021HigherOrder,Etienne} in the theoretical literature.

%\section{Near-neighbour exchange model}

\textit{Near-neighbour exchange model.---}%
We consider Ising spins on the $X$ sites of a cubic perovskite ($ABX_3$) lattice, which form corner-sharing octahedra. 
The only crystallographically allowed Ising axis connects the octahedron centres, i.e., each spin points out of one octahedron into the other.
The closest analogue of nearest-neighbour spin ice in this geometry is a twenty-vertex Ising model in which all three-in-three-out (3I3O) octahedra have equal and minimal energy:
\begin{align}
    H &= \frac{J}2 \sum_{\mathrm{octah.}\,o} \Big(\sum_{i\in o} \vec\sigma_i \cdot \hat e_\mathrm{out}\Big)^2,
    \label{eq: ideal J1-J2 Hamiltonian}
\end{align}
where $\hat e_\mathrm{out}$ is the unit vector pointing out of octahedron $o$ at site $i$ and $\vec\sigma_i$ is constrained to be $\pm \hat e_\mathrm{out}$.
Since nearest-neighbour Ising spins are perpendicular, Heisenberg interactions between them would be inactive. Therefore, realising~\eqref{eq: ideal J1-J2 Hamiltonian} requires either Dzyaloshinskii--Moriya (DM) or symmetric off-diagonal exchange of equal strength to the ferromagnetic across-octahedron interaction ($J=J_1=J_2$, cf. \cref{fig: ideal ice pinch points}a):
\begin{align}
    H &= J_1 \sum_{\langle ij\rangle} \hat{d}_{ij}\cdot (\vec\sigma_i\times\vec\sigma_j) - J_2 \sum_{\langle\langle i\leftrightarrow j\rangle\rangle} \vec\sigma_i\cdot \vec\sigma_j + \mathrm{const.};
    \label{eq: general J1-J2 Hamiltonian}
\end{align}
the DM vector $\hat d_{ij} = \hat e_{\mathrm{out},i} \times \hat e_{\mathrm{out},j}$ is the only one allowed by lattice symmetry~\cite{Moriya1960}.

\begin{figure}
    \centering
    \includegraphics{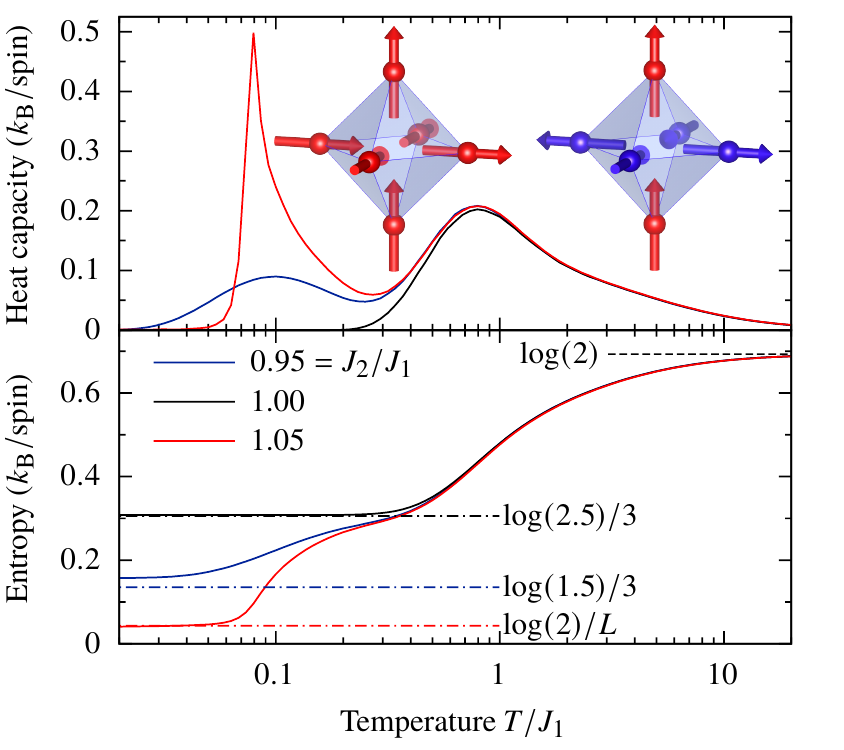}
    \caption{Heat capacity (top) and entropy (bottom) of~\eqref{eq: general J1-J2 Hamiltonian} for three values of $J_2/J_1$. Inset: a ferromagnetic (left) and a dipolar-quadrupolar (right) octahedron; red spins are in dipolar, blue in quadrupolar alignment.}
    \label{fig: thermo exchange}
\end{figure}

We computed thermodynamic properties of \cref{eq: ideal J1-J2 Hamiltonian} on a system of $64^3$ unit cells using an efficient cluster Monte Carlo algorithm adapted from Ref.~\cite{Otsuka2014ClusterIce} and plotted heat capacity and entropy as a function of temperature in \cref{fig: thermo exchange}. The heat capacity shows a single Schottky peak at $T\sim J$, corresponding to the proliferation of non-3I3O octahedra, viz.\ emergent monopoles.
This Schottky peak does not account for the full high-temperature entropy of the system; the numerically obtained residual entropy is in good agreement with the Pauling estimate~\cite{Pauling1935TheArrangement} for the twenty-vertex model, $S_0 = k_\mathrm{B}\log(2.5)/3\approx 0.305k_\mathrm{B}$ per spin.

Similar to pyrochlore ice, long-wavelength properties of this ground-state manifold can be understood by coarse-graining the spins $\vec\sigma$ into a polarisation field $\vec P$.
The free energy density of this field is purely entropic in origin and, in the first approximation, proportional to $|\vec P|^2$. 
Together with the divergence-free constraint $\nabla\cdot \vec P = 0$ imposed by the 3I3O rule, this gives rise to long-range dipolar correlations~\cite{Henley2005Power-lawAntiferromagnets,Isakov2004DipolarMagnets,Henley2010TheSystems}, which can be observed as sharp pinch points in the magnetic structure factor at reciprocal lattice vectors (\cref{fig: ideal ice pinch points}b). 
Non-3I3O defects are sources of the $\vec P$ field and, in dipolar systems, act as emergent monopoles of the physical magnetic field~\cite{Castelnovo2008MagneticIce}.

The above Coulomb-phase description, however, only holds for the fine-tuned point~\eqref{eq: ideal J1-J2 Hamiltonian}.
For $J_2>J_1$, ferromagnetic alignment between spins across an octahedron is favoured.
Such alignment is compatible with the 3I3O rule: namely, one spin points in and one out along each body diagonal of the octahedron (see inset of \cref{fig: thermo exchange}).
This alignment across each octahedron results in ground states with ferromagnetically aligned chains along all $\langle100\rangle$ directions.
These chains are, however, decoupled, leading to uniform planes of diffuse neutron scattering at $\{hkL\} (L\in \mathbb{Z})$ type wave vectors~\cite{Overy2016StructuralIce,Comes1970PerovskitesFerroelectriques} and a subextensive ground-state entropy of $k_\mathrm{B} \log 2 \times 3L^2$ per $L^3$ unit cells:
this entropy is indeed found in loop-update~\cite{Melko2004MonteModel} Monte Carlo simulations of~\eqref{eq: general J1-J2 Hamiltonian} with $L=16$ (\cref{fig: thermo exchange}).
The heat capacity peak at $T\sim (J_2-J_1)$ suggests that the chains align in a Kasteleyn transition resembling that of pyrochlore spin ice in a $[100]$ field~\cite{Jaubert2008Kasteleyn}.

$J_2 < J_1$, by contrast, favours 3I3O octahedra that are as antiferromagnetic as possible.
However, if all three pairs of opposite spins are antiferromagnetically aligned (i.e., both point in or both out), the total number of spins pointing into the octahedron is even, which frustrates the 3I3O rule.
For $0 < J_2< J_1$, the optimal octahedra are still 3I3O, but not fully antiferromagnetic, pairs across the octahedron being two-in, two-out, and one-in-one-out.
There are 12 such \textit{dipolar-quadrupolar} octahedra, so the ground-state entropy remains extensive: Pauling's estimate yields $S_0' =k_\mathrm{B}\log(1.5)/3 \approx0.135 k_\mathrm{B}$ per spin, which is in good agreement with Monte Carlo simulations (\cref{fig: thermo exchange}).
Entropy is released in two Schottky peaks: at $T\sim J_{1,2}$, the high-temperature paramagnet crosses over into twenty-vertex spin ice, while at $T\sim (J_1-J_2)$, the twelve-vertex ground-state manifold is selected out of the latter.
Dipolar components in this manifold still form a Coulomb phase: pinch points remain sharp but their intensity is reduced due to the smaller average dipole moment of dipolar-quadrupolar octahedra (\cref{fig: ideal ice pinch points}d).
The antiferromagnetic quadrupolar components, on the other hand, enhance the magnetic structure factor in rods at $(1/2,1/2,\ell)$ and similar wave vectors without new singular features.
Such fragmentation \textit{within} the ice manifold has no analogue on the pyrochlore lattice,
and, as quadrupoles do not contribute to the coarse-grained field $\vec P$, it is not fully captured by Coulomb-phase theory.

%\section{Dipolar spin ice}

\textit{Dipolar spin ice.---}%
Next, we consider adding dipolar interactions between the large rare-earth Ising moments:
\begin{align}
    H &= D\ell^3  \sum_{ij}\left[\frac{\vec\sigma_i \cdot \vec\sigma_j }{r_{ij}^3} - \frac{3(\vec\sigma_i \cdot \vec{r}_{ij}) (\vec\sigma_j \cdot \vec{r}_{ij})}{r_{ij}^5}\right] \nonumber\\*
    & \qquad{} 
    - J_2 \sum_{\langle\langle i\leftrightarrow j\rangle\rangle} \vec\sigma_i \cdot \vec\sigma_j 
    - J_2' \sum_{\langle\langle i j\rangle\rangle'} \vec\sigma_i \cdot \vec\sigma_j,
    \label{eq: dipolar Hamiltonian}
\end{align}
where $\ell$ is the distance of nearest-neighbour spins and $D=\mu_0\mu^2/(4\pi\ell^3)$ is the dipolar interaction-energy scale.
For simplicity, we set $J_1=0$; indeed, adding multiples of~\eqref{eq: ideal J1-J2 Hamiltonian} to the Hamiltonian only renormalises the energy cost of monopoles, so low-temperature effects of $J_1$ and $J_2$ are equivalent.
On the other hand, we add second-neighbour interactions between spins in different octahedra ($J_2'$ in \cref{fig: ideal ice pinch points}a) as a representative of all subleading exchange couplings; in antiperovskite materials, we expect $J_2'\ll J_2$ as there is no obvious exchange pathway between $2'$ neighbours.

\begin{figure}
    \centering
    \includegraphics{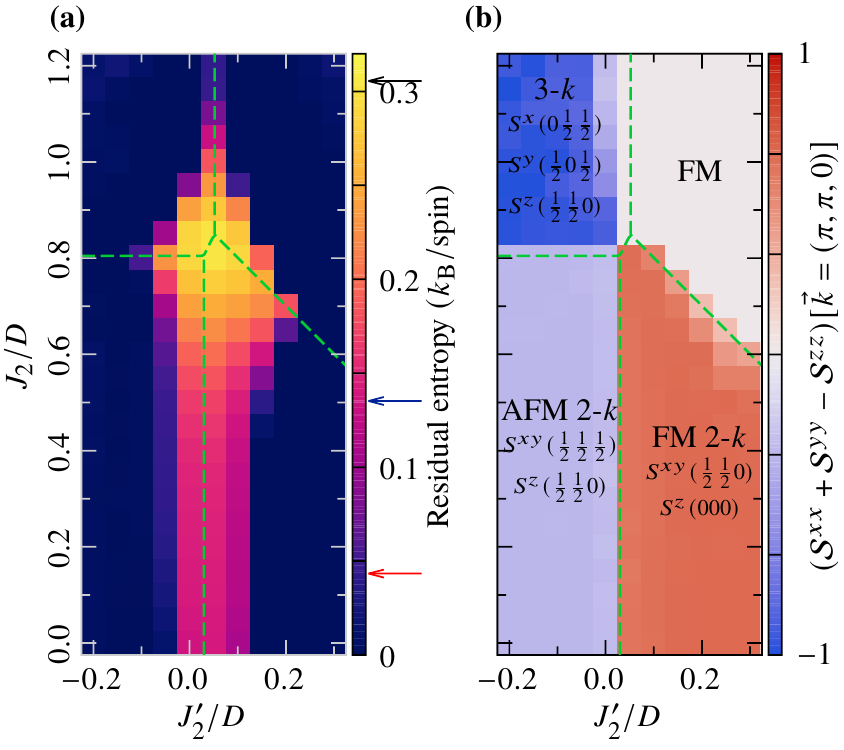}
    \caption{(a) Residual entropy at $T=0.25D$ (below spin freezing) in single-spin-flip dynamics. The coloured arrows indicate the Pauling entropy estimates of the near-neighbour model~\eqref{eq: general J1-J2 Hamiltonian}.
    (b) Composite order parameter of the four ordered phases at $T=0.1D$ in loop-update dynamics, which remains in equilibrium down to this temperature.}
    \label{fig: thermo dipolar}
\end{figure}

On the pyrochlore lattice, dipolar interactions can be decomposed into a leading term that leaves the monopole-free sector exactly degenerate and residual interactions that decay as $r^{-5}$ and are thus negligible beyond nearest neighbours~\cite{Isakov2005WhyRules}: 
that is, the two-in-two-out manifold in dipolar pyrochlore ice remains degenerate to a good approximation.
A similar decomposition of the dipolar Hamiltonian is possible on the antiperovskite lattice as well~\cite{supplement};
however, as the nearest-neighbour residual interaction is much stronger than that across an octahedron, a purely dipolar Hamiltonian would be projectively equivalent to the twelve-vertex model found above for $J_2<J_1$~\cite{supplement}.
Emulating the twenty-vertex Coulomb phase requires compensating the residual terms with a ferromagnetic $J_2^*\approx0.75D$~\cite{supplement};
indeed, the entire 3I3O manifold is approximately degenerate at this point, evidenced by the confluence of all ordered phases in \cref{fig: thermo dipolar}b.

Using single-spin-flip Monte Carlo dynamics~\cite{denHertog2000DipolarMagnets}, we indeed find an extensive residual entropy for $J_2\lesssim J_2^*, J_2'\approx0$.
At $J_2\approx J_2^*$, the residual entropy is consistent with that of the twenty-vertex model;
as $J_2$ is lowered, this value decreases continuously to the residual entropy of the twelve-vertex fragmented spin ice of dipolar-quadrupolar octahedra (\cref{fig: thermo dipolar}a).
These icelike phases are stabilised by the dipolar interactions, which cause the single-spin-flip dynamics (realised in materials by thermal fluctuations) to become glassy~\cite{supplement}, thus preventing any low-temperature ordering transition, similar to dipolar pyrochlore ice~\cite{Melko2004MonteModel,Snyder2004Low-temperatureIce,Jaubert2009SignatureIce,Jaubert2011MonopoleDynamics,Tomasello2019SlowDynamics,Jonathan}.
Indeed, simulations using loop-update dynamics find ordered ground states with zero residual entropy for all values of $J_2,J_2'$ (\cref{fig: thermo dipolar}b).

\begin{figure}
    \centering
    \includegraphics{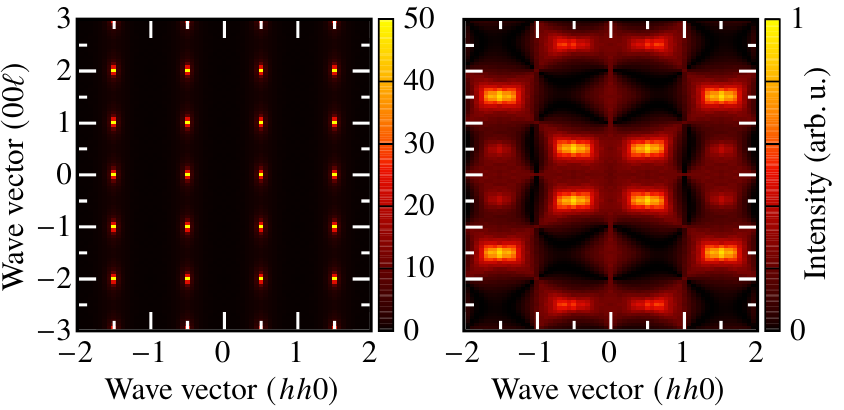}
    \caption{Polarised neutron-scattering pattern of~\eqref{eq: dipolar Hamiltonian} at $J_2=0.4D$, $J_2'=0.05 D$. Neutrons are polarised along $[1\bar10]$. The non-spin-flip channel (left) is strongly peaked at $(1/2,1/2,0)$, with some visible broadening along the $q_z$ direction. The spin-flip channel (right) remains diffuse, with pinch points at $\Gamma$ and smooth intensity maxima at $(1/2,1/2,1/2)$.}
    \label{fig: dipolar pinch points}
\end{figure}

Magnetic structure factors in the high-residual-entropy regime resemble closely those of the $J_1$--$J_2$ model above. 
Neutron-scattering patterns at $J_2\approx J_2^*$ are similar to those of~\eqref{eq: ideal J1-J2 Hamiltonian}~\cite{supplement},
while the $J_2 < J_2^*$ case (\cref{fig: dipolar pinch points}) resembles the twelve-vertex model (\cref{fig: ideal ice pinch points}c,d), with two important differences.
First, dipolar pinch points have higher intensity, while the broad quadrupolar maxima at $(1/2,1/2,1/2)$ are weaker.
Second, sharp peaks appear at $(1/2,1/2,0)$ due to ordering of the quadrupolar components of the octahedra; 
however, the ordered moment remains quite small, as evidenced by a substantial diffuse component surrounding the peaks.
Both effects are due to the low-temperature glassiness of single-spin-flip dynamics, which sabotages both the full suppression of ferromagnetic octahedra and the full development of order.

%\section{Ordered phases}

\textit{Ordered phases.---}%
Such spin freezing can effectively be lifted in Monte Carlo simulations by adding short-loop updates that respect the 3I3O ice rules~\cite{Melko2004MonteModel}. 
These reveal four low-temperature ordered phases that all meet near the optimally compensated spin-ice point~(\cref{fig: thermo dipolar}b). 
At $J_2\gtrsim J_2^*$, ferromagnetic octahedra are favourable, resulting in aligned chains of spins, similar to the proposed ground states of pyrochlore ice~\cite{Melko2004MonteModel,McClarty2015Chain,Henelius2016Chain}.
If $J_2'\gtrsim0$, these chains align into a three-dimensional canted ferromagnet; 
for antiferromagnetic $J_2'$, each sublattice orders along the $\{1/2,1/2,0\}$ wave vector normal to its spin direction, resulting in a three-$k$ structure with full cubic symmetry.
Both transitions are strongly first-order~\cite{supplement}, similar to dipolar pyrochlore ice~\cite{Melko2004MonteModel}.

\begin{figure}
    \centering
    \includegraphics{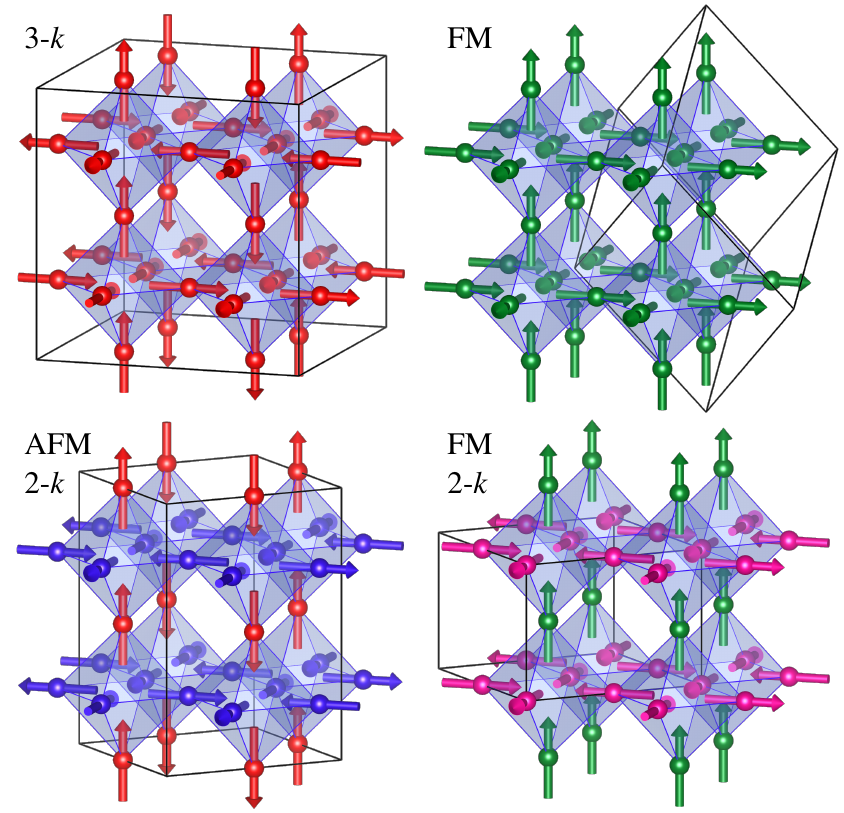}
    \caption{ Magnetic structures of the four ordered phases in Fig.~\ref{fig: thermo dipolar}b. 
    Colours indicate inequivalent ordering wave vectors: green---$(000)$; blue---$(1/2,1/2,1/2)$; red---$(1/2,1/2,0)$ normal to spin direction; purple---$(1/2,1/2,0)$ in the plane of spin directions. 
    The black boxes indicate the magnetic unit cell of each order.}
    \label{fig: structures}
\end{figure}

On the other hand, phases at $J_2\lesssim J_2^*$ are composed of dipolar-quadrupolar octahedra:
the two components order independently, which breaks cubic rotational symmetry and results in two-$k$ structures.
Ordering wave vectors are again controlled by the sign of $J_2'$, as shown in \cref{fig: structures};
neutron-scattering signatures of the ordered phases are discusssed in the supplement~\cite{supplement}.
As the ordered state is described by two independent order parameters with distinct symmetries, the ordering transition occurs in two stages:
first, quadrupolar components order in a continuous transition, followed by the dipolar ones at a lower temperature, which leads to a jump in both order parameters~\cite{supplement}.
Between these transitions, we anticipate another kind of fragmented spin ice, in which the quadrupolar components show partial ordering while dipolar ones remain in a Coulomb phase.

%\section{Conclusion}

\textit{Conclusion.---}%
In summary, we have studied dipolar and purely near-neighbour exchange Ising models on a lattice of corner-sharing octahedra, as may be realised in rare-earth antiperovskite materials.
Similar to spin ice on the pyrochlore lattice, the phase diagram of these systems can be understood in terms of a twenty-vertex model of three-in-three-out octahedra, which gives rise to a low-temperature Coulomb phase with extensive ground-state entropy, pinch-point correlations, and emergent magnetic monopoles.
While such a model is fine-tuned, glassiness induced by dipolar interactions in spin ice stabilises the Coulomb phase for a finite range of exchange parameters.
More interestingly, perturbing away from this fine-tuned point results in ``fragmented spin ice,'' where spins around each octahedron exhibit both dipolar and quadrupolar alignment;
while the former remain in a Coulomb phase, the latter also contribute distinctive features to structure factors.
Such a fragmented phase is not possible in pyrochlore ice (except in strained thin films~\cite{Jaubert2017ThinFilm,Bovo2019FModel}), as all 2I2O tetrahedra are symmetry-equivalent to one another; 
by contrast, it is generic on the antiperovskite lattice as it is stabilised by the dipolar interaction for a wide and experimentally relevant range of parameters.
Further-neighbour interactions break the approximate degeneracy of these ice manifolds, resulting in a variety of ordered phases.
In particular, perturbing the fragmented spin ice gives rise to two-$k$ structures 
as the dipolar and quadrupolar components give rise to symmetry-inequivalent order parameters at different propagation vectors and thus two distinct ordering transitions at different temperatures.

Our results demonstrate that both mechanisms by which dipolar interactions stabilise pyrochlore ice, the projective equivalence of dipolar and near-neighbour exchange models~\cite{denHertog2000DipolarMagnets,Isakov2005WhyRules} and glassy low-temperature spin dynamics~\cite{Melko2004MonteModel,Snyder2004Low-temperatureIce,Jaubert2009SignatureIce}, remain active on other lattices and help stabilise analogous Coulomb phases. 
This is critical for antiperovskite materials, as the projective-equivalence mechanism is insensitive to low-temperature distortions of the crystal structure~\cite{Glazer1972PerovskiteTilting},
and may also help stabilise more complex models, several of which exhibit fractonic features~\cite{Benton2021HigherOrder,Etienne}.

In fact, non-pyrochlore lattices may exhibit richer physics, precisely on account of the several interaction terms that need to be fine-tuned to realise the perfect Coulomb phase. A case in point is the antiperovskite lattice, where frustration between antiferromagnetic correlations and the 3I3O rule results in fragmented spin ice, without direct analogues in bulk pyrochlores.
This opens up new experimental directions for frustrated magnetism: indeed, certain rare-earth antiperovskites are known to avoid magnetic ordering down to very low temperatures~\cite{Hoppe2002GdAntiperovskite,Kloss2017HoAntiperovskite}, which may indicate spin-ice physics.
On the theoretical side, developing field theories for the fragmented ice phase promises considerable challenges due to the frustration needed to bring it about. (The long-wavelength physics of Coulomb phases is essentially unfrustrated~\cite{Isakov2004DipolarMagnets,Henley2005Power-lawAntiferromagnets}.) The coexistence of dipolar and quadrupolar features also raises the possibility of a higher-order gauge theory description, with potentially exciting consequences for both excitations and ordering transitions.

%\section{Acknowledgements}

\begin{acknowledgments}
The data underlying this work and the code used to generate it are available on Zenodo~\cite{Data,Code}.

We thank Claudio Castelnovo, Dmitry Khalyavin, and Étienne Lantagne-Hurtubise for useful discussions.
Figs.~\ref{fig: ideal ice pinch points}a and~\ref{fig: structures} were made using VESTA~\cite{Momma2011VESTAData}.
All heat maps use perceptionally uniform colour maps developed in Ref.~\cite{Kovesi2015GoodThem}.
A.~Sz.\ gratefully acknowledges the ISIS Neutron and Muon Source and the Oxford--ShanghaiTech collaboration for support of the Keeley--Rutherford fellowship at Wadham College, Oxford.
Computing resources were provided by STFC Scientific Computing Department’s SCARF cluster.
For the purpose of open access, the authors have applied a Creative Commons Attribution (CC-BY) licence to any author accepted manuscript version arising.
\end{acknowledgments}

\bibliography{references,paper}

\end{document}

\title{Supplementary material to\\``Fragmented spin ice and multi-$k$ ordering in rare-earth antiperovskites''}

\author{Attila Szab\'{o}}
\address{Rudolf Peierls Centre for Theoretical Physics, University of Oxford, Oxford OX1 3PU, UK}
\address{ISIS Facility, Rutherford Appleton Laboratory, Harwell Campus, Didcot OX11 0QX, UK}
\author{Fabio Orlandi}
\address{ISIS Facility, Rutherford Appleton Laboratory, Harwell Campus, Didcot OX11 0QX, UK}
\author{Pascal Manuel}
\address{ISIS Facility, Rutherford Appleton Laboratory, Harwell Campus, Didcot OX11 0QX, UK}

\date{\today}

\maketitle

\section{Details of Monte Carlo simulations}
\label{ssec: Monte Carlo}

We performed Monte Carlo simulations using three complementary methods:
\begin{itemize}
    \item Single-spin-flip Metropolis algorithm. In addition to equilibrium thermodynamic properties, this method emulates the dynamics of the experimental system under random thermal fluctuations, making it appropriate for studying, e.g., glassiness~\cite{Jaubert2009SignatureIce,Jaubert2011MonopoleDynamics}.
    \item Short-loop-update Metropolis algorithm~\cite{Melko2004MonteModel}. Since these updates do not change monopole number on any octahedron, we combine them with single-spin-flip updates.
    \item Loop-string algorithm~\cite{Otsuka2014ClusterIce} (see~\cref{ssec: otsuka})  for the idealised twenty-vertex model
    \begin{equation}
        H = \frac{J}2 \sum_{\mathrm{octah.}\,o} \Big(\sum_{i\in o} \vec\sigma_i \cdot \hat e_\mathrm{out}\Big)^2.
        \label{seq: ideal ice}
    \end{equation}
\end{itemize}
We use simulated annealing for all simulations: after obtaining a number of samples, we reduce the temperature by a constant factor (specified as an integer number of $T$-points per decade). At each $T$-point, we discard a number of ``burn-in'' samples first, and record subsequent ones without regard to autocorrelation times. We perform several rounds of this procedure to ensure some independent samples at all temperatures. 
Specific details of all simulations are summarised in \cref{stab: Monte Carlo}.

All simulations are done on cube-shaped simulation boxes ($L=16$ for all simulations except for those with the loop-string algorithm, where $L=64$) with periodic boundary conditions. Dipolar interactions are handled using Ewald summation~\cite{deLeeuw1980SimulationConstants}; surface terms are omitted as they force ferromagnetic orders to break up into two large domains, which makes interpreting the data significantly harder (cf. footnote~\ref{sfootnote: ewald} on page~\pageref{sfootnote: ewald}).

In all plots of magnetic structure factors and related order parameters, the structure factors obtained from the Monte Carlo simulation are symmetrised with respect to all point-group symmetries of the cubic antiperovskite lattice.
Heat capacities are obtained from the formula $C_V = \var E/ k_\mathrm{B} T^2$. Instead of directly integrating $C_V/T$, entropy is calculated by summing the differences $\Delta S = \Delta E/\overline{T}$ between neighbouring $T$-points in the simulation: this approach is much more robust near phase transitions.

\begin{table}
    \centering
    \setlength{\tabcolsep}{1.25ex}
    \begin{tabular}{cccccccc} \hline\hline
        Model & Algorithm & $T_\mathrm{max}$ & $T_\mathrm{min}$ & $T$-points & \#burn-in & \#samples & \#rounds \\\hline
        \eqref{seq: ideal ice} & loop-string & $100J$ & $0.1J$ & 30/decade & 64 & 8192 & 1 \\
        $J_1$--$J_2$ exchange & short-loop-update & $20J_1$ & $0.01J_1$ & 30/decade & 256 & 4096 & 256 \\
        dipolar & single-spin-flip & $1D$ & $0.25D$ & 40/decade & 512 & 4096 & 48 \\
        dipolar & short-loop-update & $20D$ & $0.1D$ & 30/decade & 128 & 512 & 48 \\\hline
    \end{tabular}
    \caption{Details of all Monte Carlo simulations shown in the main text. All single-spin-flip simulations remain in equilibrium at $T\ge D$, so results from the loop-update simulations were used as equivalent. }
    \label{stab: Monte Carlo}
\end{table}

\subsection{Loop-string algorithm for the twenty-vertex model}
\label{ssec: otsuka}

In the following, we summarise our adaptation of the loop-string algorithm of Ref.~\cite{Otsuka2014ClusterIce} to the antiperovskite lattice.

The algorithm makes use of loop-string graphs: in these, each spin is connected to at most one other in each of the two octahedra it belongs to (i.e., edges may connect nearest neighbours or spins across an octahedron). A graph $G$ is said to be compatible with a spin configuration $S$ (written as $S\sim G$) if one of the spins at the ends of each edge of $G$ points into the shared octahedron, while the other points out: then, spins along all loops/strings in the graph will be aligned start to end, so flipping a string may only change the monopole charge at its two ends. Now, if we can assign positive weights $W(G)$ to all graphs such that the Boltzmann factor of $S$ is given the total weight of all graphs it is compatible with,
\begin{equation}
    W(S) \equiv e^{-\beta H(S)} = \sum_{G\sim S} W(G),
    \label{seq: Otsuka graph condition}
\end{equation}
we can generate spin configurations according to the Boltzmann distribution using the following algorithm:
\begin{enumerate}
    \item Given a spin configuration $S$, choose a graph $G\sim S$ with probability $W(G)/W(S)$.
    \item Flip all loops/strings of $G$ with probability $1/2$ to obtain the new configuration $S'$.
\end{enumerate}
Indeed, the probability of moving from spin configuration $S$ to $S'$ is
\begin{equation}
    P(S\to S') = \sum_{S\sim G \sim S'} \frac{W(G)}{W(S)} \frac1{2^{n(G)}},
\end{equation}
where $n(G)$ is the number of loops/strings in $G$: this clearly satisfies the detailed balance condition $P(S\to S')W(S) = P(S'\to S)W(S')$.
The algorithm is extremely efficient at generating uncorrelated samples as there is no rejection and the autocorrelation function $\sigma_i(t)\sigma_i(t+\Delta t)$ vanishes for any $\Delta t > 0$.

Ref.~\cite{Otsuka2014ClusterIce} gives an explicit construction of $W(G)$ for nearest-neighbour pyrochlore ice, exploiting the fact that their Boltzmann factor can be written as a product of single-tetrahedron terms: $W(S) = \prod_i w(s_i)$, where $w(s_i)$ only depends on the spin configuration of tetrahedron $i$.
Likewise, graphs $G$ can be split up into edge layouts $g_i$ on each tetrahedron that can be combined arbitrarily.
Finally, the condition $G\sim S$ can also be checked tetrahedron by tetrahedron: we write $g\sim s$ to mean that all edges in $g$ connect an incoming and an outgoing spin in $s$.
Therefore, if we can assign positive weights $w(g)$ to these edge layouts such that 
\begin{equation}
    w(s) = \sum_{g\sim s} w(g),
    \label{seq: Otsuka local condition}
\end{equation}
setting $W(G) = \prod_i w(g_i)$ will satisfy~\cref{seq: Otsuka graph condition}:
\begin{equation}
    W(S) = \prod_i w(s_i) = \prod_i \bigg(\sum_{g_i\sim s_i} w(g_i) \bigg) = \sum_{\{g_i\}\sim\{s_i\}} \prod_i w(g_i) = \sum_{G\sim S} W(G).
\end{equation}

Since the idealised octahedral spin-ice Hamiltonian~\eqref{seq: ideal ice} is also a sum of terms that act on single octahedra, the above construction is straightforward to generalise to it as long as $w(g)$ satisfying~\eqref{seq: Otsuka local condition} can be found.
Since the local Boltzmann factor $w(s)$ only depends on the number of edges pointing into/out of the octahedron, we try to find $w(g)$ that only depend on the number of ``loose ends'' of strings ending on the octahedron: $w(g) = w_n$ with $n=0,2,4,6$. \cref{seq: Otsuka local condition} becomes
\begin{equation}
    \begin{aligned}
        w(\mathrm{3I3O}) = 1 &= w_6 + 9 w_4 + 18 w_2 + 6 w_3 \\
        w(\mathrm{4I2O}) = e^{-2\beta J} &= w_6 + 8 w_4 + 12 w_2 \\
        w(\mathrm{5I1O}) = e^{-8\beta J} &= w_6 + 5w_4 \\
        w(\mathrm{6I}) = e^{-18\beta J} &= w_6
    \end{aligned}
    \implies
    \begin{aligned}
        w_0 &= (10-15e^{-2\beta J} + 6e^{-8\beta J}-e^{-18\beta J})/{60}\\
        w_2 &= (5e^{-2\beta J}- 8e^{-8\beta J} + 3e^{-18\beta J})/{60}\\
        w_4 &= (e^{-8\beta J}-e^{-18\beta J})/5\\
        w_6 &= e^{-18\beta J}
    \end{aligned}
    \label{eq: Otsuka weights}
\end{equation}
One can verify that these $w$ are positive for all $\beta>0$. The full algorithm can thus be written as follows:
\begin{enumerate}
    \item For all octahedra, do:
    \begin{enumerate}
        \item Choose the number $n=0,2,4,6$ of ``loose ends'' to leave from the probability distribution $M_{n}(s) w_n / w(s)$, where $M_n(s)$ are the integer coefficients appearing in~\eqref{eq: Otsuka weights}.
        \item Choose $(6-n)/2$ incoming and outgoing spins uniformly at random and connect them pairwise with edges.
    \end{enumerate}
    \item Flip each loop/string in the resulting graph with probability $1/2$.
\end{enumerate}

We note that Ref.~\cite{Otsuka2015Loop-stringIce} introduced a loop-string algorithm, based on the above one, for dipolar spin ice. However, given the mixing efficiency of short-loop updates and the added complications and worse vectorisation properties of this algorithm, we decided against adapting it to antiperovskites.
The approach also does not generalise to dipolar-quadrupolar ice (that is, $0<J_2<J_1$, $T\ll J_1-J_2$), as there is no choice of positive $w(g)$ that satisfy~\eqref{seq: Otsuka local condition}.

\clearpage
\subsection{Autocorrelation times}

\begin{figure}
    \centering
    \includegraphics{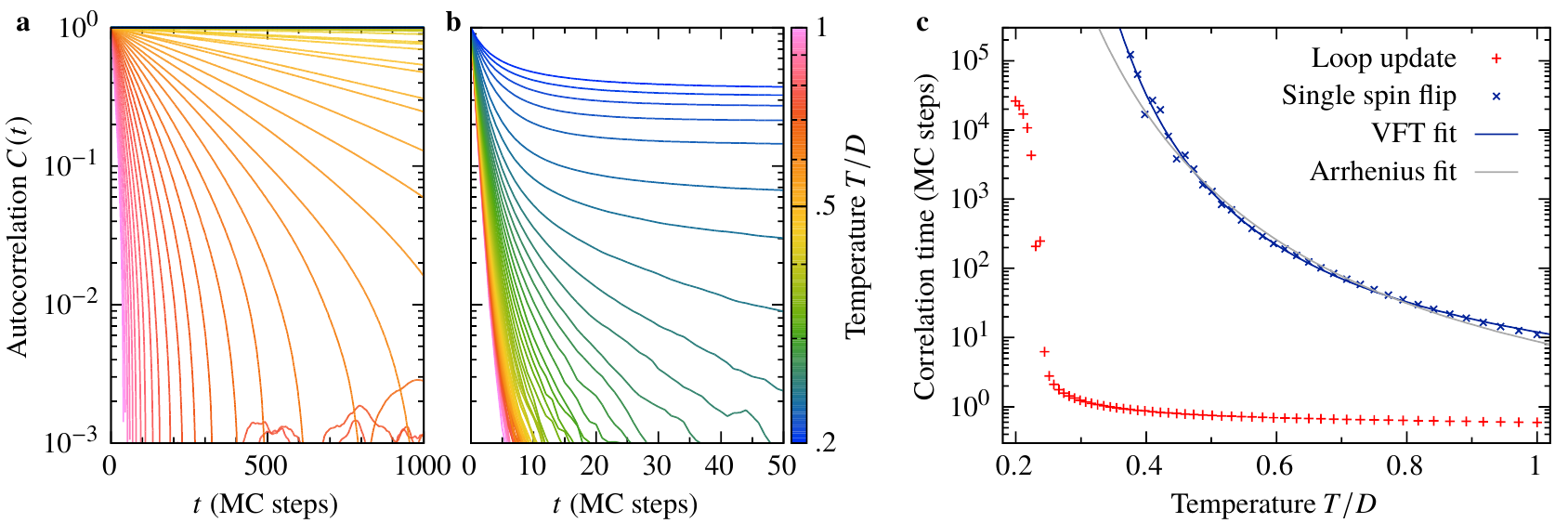}
    \caption{(a,b) Log-linear plot of the autocorrelation function $C(t)$ for single-spin-flip (a) and loop-update (b) dynamics. Initially, all curves decay exponentially, allowing an autocorrelation time $\tau(T)$ to be extracted. (c) Autocorrelation time $\tau$ as a function of temperature. Below the ordering transition at $T\approx0.23D$, $\tau$ of the loop-update dynamics diverges; the finite values are an artefact of automatically fitting a decaying exponential to the $C(t)$ curve.}
    \label{sfig: autocorr}
\end{figure}

We performed $t_\mathrm{tot}=16\,384$ iterations%
\footnote{For single-spin-flip dynamics, one ``iteration'' consists of attempting to flip all spins sequentially. In loop-update dynamics, this is followed by attempting to flip a number of randomly chosen loops until the total number of spins in these loops first exceeds the total number of spins.} 
of both the loop-update and the single-spin-flip Monte Carlo algorithms for $16^3$ unit cells of the dipolar Hamiltonian with exchange constants $J_2=0.8D$, $J_2'=0.05D$ using 80 $T$-points per decade between $T=D$ and $T=0.2D$.
The full spin configuration after each iteration was stored and used to evaluate the spin autocorrelation function
\begin{equation}
    C(\Delta t) = \frac1{N_\mathrm{spin}t_\mathrm{max}} \sum_{t=0}^{t_\mathrm{max}} \sum_i \sigma_i(t) \sigma_i(t+\Delta t)
\end{equation}
up to $\Delta_\mathrm{max}=2048$ time steps ($t_\mathrm{max}=t_\mathrm{tot}-\Delta_\mathrm{max}$). These show exponential decay at all temperatures (\cref{sfig: autocorr}a,b), which allows us to extract the autocorrelation time accurately by fitting $C(t) \sim e^{-t/\tau}$. These are plotted for both algortithm in \cref{sfig: autocorr}c.
Loop-update dynamics mixes rapidly at all temperatures down to an ordering transition at $T\approx0.23D$, below which $C(t)$ does not tend to 0 at any time;
by contrast, single-spin dynamics slows down continuously with temperature.
The autocorrelation time fits the Vogel--Fulcher--Tammann relation $\tau\sim e^{-\Delta/(T-T_0)}$ significantly better than the simple activated form $\tau\sim e^{-\Delta/T}$.
This indicates a glass transition just above $T_0\approx0.15D$;
while this may be below the ordering transition temperature, the slow intrinsic dynamics of Ising-like rare-earth ions~\cite{Jaubert2009SignatureIce,Jaubert2011MonopoleDynamics,Tomasello2019SlowDynamics} and the exponential increase in autocorrelation times result in experimentally unfeasible time scales for observing the ordered phase.

\clearpage
\section{Projective equivalence}
\label{ssec: projective equivalence}

\begin{figure}
    \centering
    \includegraphics{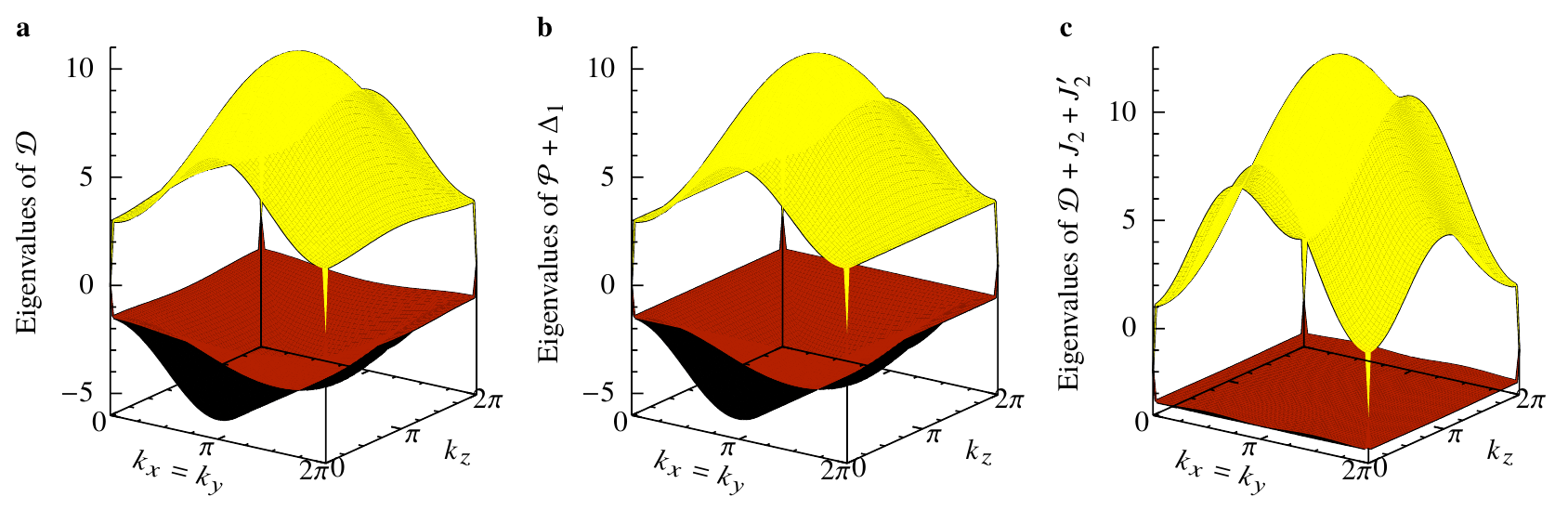}
    \caption{Dispersion of spin coupling energies in the $(hhk)$ plane. (a) Dipolar interactions $\mathcal D$ on the antiperovskite lattice, computed using Ewald summation on $128^3$ unit cells. (b) Model dipolar interactions $\mathcal P$ with first-neighbour residual interaction $\Delta_1$, visually indistinguishable from the spectrum of $\mathcal{D}$. (c) Dipolar interactions with ferromagnetic coupling $J_2=0.85 D$ across octahedra and $J_2'=0.05D$ along $2'$ pathways: the lower two bands look perfectly flat and degenerate at this scale.}
    \label{sfig: spectrum}
\end{figure}

Due to translation invariance, we can write the dipolar interactions between spins on the antiperovskite lattice as
\begin{equation}
    H_D = \frac{D}2 \sum_{\mu,\nu,\vec r_\mu,\vec r_\nu} \sigma_\mu(\vec r_\mu) \mathcal{D}_{\mu\nu}(\vec r_\mu-\vec r_\nu) \sigma_\nu(\vec r_\nu),
    \label{seq: dipolar Hamiltonian trans inv}
\end{equation}
where $\mu,\nu=x,y,z$ index the three sublattices, $\vec r_{\mu,\nu}$ run over sites within the respective sublattices, and
\begin{equation}
    \mathcal{D}_{\mu\nu}(\vec r) = \frac{\hat e_\mu\cdot \hat e_\nu }{r^3} - \frac{3(\hat e_\mu \cdot \vec{r}) (\hat e_\nu \cdot \vec{r})}{r^5},
\end{equation}
assuming $\vec r$ is measured in units of the nearest-neighbour distance $\ell$. \cref{seq: dipolar Hamiltonian trans inv} can be Fourier transformed as
\begin{align}
    H_D &= \frac{D}{2N} \sum_{\mu,\nu,k} \mathcal{D}_{\mu\nu}(\vec k) \sigma_\mu(-\vec k)  \sigma_\nu(\vec k); &
    \mathcal{D}_{\mu\nu}(\vec k) &= \sum_{\vec r}^{\mu-\nu} \mathcal D_{\mu\nu}(\vec r) e^{-i\vec k \cdot\vec r}; & \sigma_\mu(\vec k) = \sum_{\vec r_\mu} \sigma_\mu(\vec r_\mu) e^{-i\vec k \cdot \vec r_\mu};
    \label{seq: dipolar Hamiltonian FT}
\end{align}
the spectrum of $\mathcal D$ for the $(hhk)$ plane is plotted in \cref{sfig: spectrum}a.
Unlike the pyrochlore lattice~\cite{Isakov2005WhyRules}, where two bands of the equivalent dispersion, corresponding to the six-vertex ground state manifold, are essentially dispersionless, all our bands are strongly dispersive. 
Nevertheless, two bands of the dispersion in \cref{sfig: spectrum}a are separated by a clear gap from the highest-energy one, which are also degenerate at $\vec k\to 0$. Indeed, up to $O(k^2)$ corrections, $\mathcal D_{\mu\nu}(\vec k) \propto -(\delta_{\mu\nu}/3 - k_\mu k_\nu/k^2)$.%
\footnote{\label{sfootnote: ewald}%
The exact point $k=0$ is anomalous because the $k_\mu k_\nu/k^2$ term becomes singular (but not divergent) there.
On an infinite lattice, $\mathcal{D}(\vec k=0)$ vanishes exactly due to symmetries of the dipolar interaction, resulting in the triply degenerate zero eigenvalues that shoot out of the respective bands in \cref{sfig: spectrum}a. This is inconvenient for numerical simulations, as the anomalously high eigenvalue at $k=0$ disables ferromagnetic ordering in favour of domain formation. However, omitting the surface term in Ewald summation~\cite{deLeeuw1980SimulationConstants} lowers this triply degenerate point to exactly $-\sqrt2\pi D/3$ without altering the dispersion at $k\neq0$, making the lower two bands smooth.}

This is because, in the long-wavelength limit $k\to0$, the lattice discretisation of spins is irrelevant, and $\mathcal D$ can be estimated from dipolar interactions of a continous magnetisation distribution:
\begin{align}
    D\mathcal D_{\mu\nu}(k\approx0) 
&= D \sum_{\vec r} \mathcal D_{\mu\nu}(\vec r) e^{-i\vec k\cdot \vec r}
\approx \frac{\mu_0\mu^2}{4\pi} \int\frac{d^3r}{a^3} \left(\frac{\hat e_\mu\cdot \hat e_\nu - 3 (\hat e_\mu\cdot\hat r)(\hat e_\nu\cdot\hat r)}{r^3} \right) e^{-i\vec k\cdot \vec r} \nonumber\\
&= \frac{\mu_0\mu^2}{a^3} \left[(\hat e_\mu\cdot \hat k)(\hat e_\nu\cdot \hat k) - \frac{\hat e_\mu\cdot \hat e_\nu}3\right] = -\sqrt2\pi D \left(\frac{\delta_{\mu\nu}}3 - \frac{k_\mu k_\nu}{k^2}\right).
\end{align}
That is, the band structure at small $k$ consists of two degenerate bands at $-\sqrt 2\pi D/3$ and a third one at $2\sqrt 2\pi D/3$, in perfect agreement with the numerical results in~\cref{sfig: spectrum}a.
Furthermore, the eigenspace of $\mathcal{D}$ corresponding to the low-energy eigenvalue, $\sigma_\mu k_\mu=0$, is monopole-free. 
That is, in perfect analogy with the pyrochlore model~\cite{Isakov2005WhyRules}, dipolar interactions in the long-wavelength limit can be approximated as a projector onto the monopole-free sector. Extending this to all wave vectors, we define the model dipolar interaction
\begin{align}
    \mathcal{P}_{\mu\nu}(\vec k) &= \sqrt2 \pi D \left(\frac{s_\mu s_\nu}{\sum_\lambda s_\lambda^2} -\frac13 \delta_{\mu\nu} \right); &
    s_\mu(\vec k) &= 2 \sin(k_\mu/2) &
    \because q(\vec k) &= \sum_\mu s_\mu(\vec k) \sigma_\mu(\vec k),
    \label{seq: model dipolar interaction}
\end{align}
where $q(\vec r)$ is the monopole charge of the octahedron centred on $\vec r$: $\cal P$ has the correct long-range behaviour and leaves the 3I3O ice manifold exactly degenerate.

Due to symmetry and analyticity constraints, the difference $\Delta_{\mu\nu} = \mathcal{D}_{\mu\nu}-\mathcal{P}_{\mu\nu}$ must decay at least as fast as $r^{-5}$ in real space~\cite{Isakov2005WhyRules}, thus it is dominated by nearest-neighbour corrections in any geometry.
In the pyrochlore case, this means that the entire 2I2O manifold remains approximately degenerate, as the degeneracy is only lifted by heavily suppressed further-neighbour $\Delta$-corrections~\cite{Isakov2005WhyRules}.
On the antiperovskite geometry, the same would require equal $\Delta$ for nearest neighbours and spins across an octahedron, which is not the case.
As a result, the 3I3O manifold, while separated from monopole states by a spectral gap, becomes strongly modulated:
as shown in \cref{sfig: spectrum}b, the nearest-neighbour correction $\Delta_1$ already accounts for almost all of this modulation.
Real-space values of $\mathcal D, \mathcal P$, and $\Delta$ for the exchange pathways we consider are summarised in~\cref{stab: interaction terms}.

\begin{table}
    \centering
    \setlength{\tabcolsep}{1.25ex}
    \begin{tabular}{ccccc}\hline\hline
        Distance & Direction & $\mathcal D$ & $\mathcal{P}$ & $\Delta$ \\\hline
        nearest neighbour & favours 3I3O & 1.5 & 0.6033158 & 0.8966842 \\
        across octahedron & ferromagnetic & $\sqrt2/2$ & 0.5486589 & 0.1584479 \\
        $2'$ second neighbour & antiferromagnetic & $\sqrt2/4$ & 0.2743294 & 0.0792239 \\\hline
    \end{tabular}
    \caption{Dipolar ($\mathcal D$) and model dipolar ($\mathcal P$) interaction terms and their difference ($\Delta$) along the three near-neighbour pathways considered in this work. $\mathcal{D}$ can be calculated exactly; $\mathcal P$ was obtained from numerical Fourier transforms of~\eqref{seq: model dipolar interaction} on a grid of $128^3$ unit cells, which matches the infinite-system values to at least nine significant figures.}
    \label{stab: interaction terms}
\end{table}

Realising the full twenty-vertex ground state manifold requires that the two lower, monopole-free bands of the energy dispersion be perfectly flat.
This can be achieved by adding exchange interactions $-\Delta$ between all pairs of spins, which turns $\mathcal{D}$ into the model dipolar interaction $\mathcal{P}$, which has three flat bands.
However, $\Delta$ beyond second neighbours is negligibly small; furthermore, adding multiples of~\eqref{seq: ideal ice} only changes the monopole chemical potential while leaving the monopole-free sector unchanged, so both $\Delta_1$ and $\Delta_2$ can be corrected for using ferromagnetic exchange of strength $J_2^*/D = \Delta_1-\Delta_2\approx 0.738$ across octahedra.
In practice, further-neighbour interactions move the point of flattest 3I3O bands to slightly higher values of $J_2$, see \cref{sfig: spectrum}c.

\subsection{Phase boundaries}

Knowing the ordered structures of the four ordered phases, we can evaluate their energy under the full Hamiltonian by substituting the ordered moments into~\eqref{seq: dipolar Hamiltonian FT} and counting satisfied and frustrated exchange couplings in each unit cell. 
The dipolar interaction tensor $\mathcal{D}(\vec k)$ was taken from numerical Fourier transforms of Ewald summed dipolar terms at all $\vec k\neq 0$;
these match the infinite-system values to at least nine significant figures on the $128^3$ lattice used.
At $\vec k = 0$, we used the exact $k\to 0$ limit of the lower two bands, $\mathcal{D}_{\mu\nu} = -\sqrt2 \pi D \delta_{\mu\nu}/3$ (note that ferromagnetically aligned spins automatically satisfy the ice rules, so they are continuations of the lower bands).
This yields the following energies per unit cell:
\begin{subequations}
\label{seq: phase boundaries}
\begin{alignat}{7}
    E_\mathrm{FM} &= && -3 J_2 && +6 J_2' && -2.22144147 D;\\
    E_{3k} &= && -3 J_2 && +6 J_2' && -2.83916315 D;\\
    E_{\textrm{FM-}2k} &= -4 J_1 && +J_2 && -2 J_2' && -5.82563485 D;\\
    E_{\textrm{AFM-}2k} &= -4 J_1 && +J_2 && +6 J_2' && -6.05897801 D.
\end{alignat}
\end{subequations}
The phase boundaries shown in Fig.~3 of the main text follow by setting pairs of these equal.

\clearpage
\section{Thermodynamics of dipolar spin ice}

\begin{figure}
    \centering
    \includegraphics{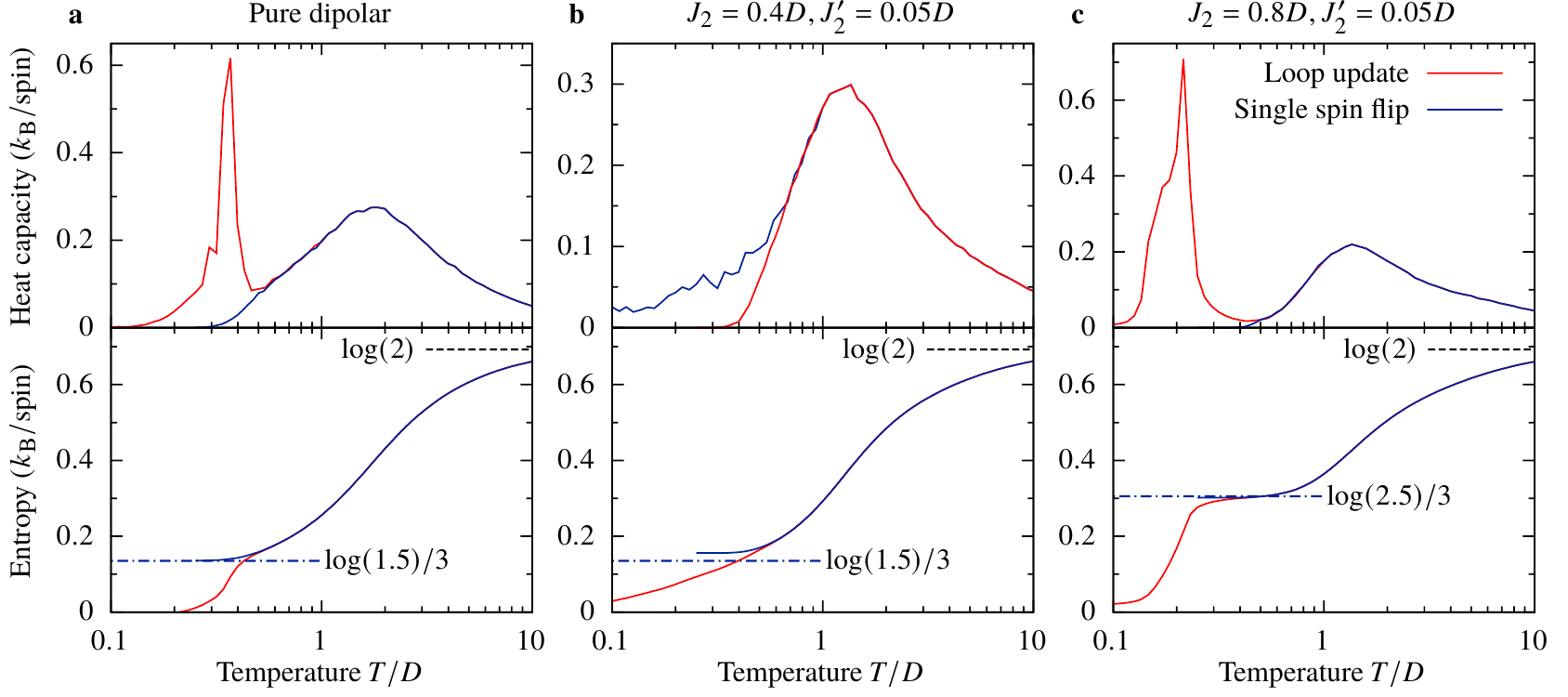}
    \caption{Heat capacity (top) and entropy (bottom) of the dipolar spin-ice model for three representative values of exchange couplings with loop-update (red) and single-spin-flip (blue) dynamics. 
    In the latter case, the heat-capacity curve consists of a single Schottky peak; the residual entropy is consistent with dipolar-quadrupolar ice for $J_2<J_2^*\approx0.8 D$ and with the idealised twenty-vertex spin ice~\eqref{seq: ideal ice} for $J_2\approx J_2^*$. 
    Loop-update dynamics always reaches an ordered state with no residual entropy.
    Depending on the low-temperature ordered phase, the residual entropy is released in qualitatively different ways; 
    by contrast, the single-spin-flip Schottky peaks in the first two panels are nearly identical.}
    \label{sfig: dipolar thermo}
\end{figure}

We calculated heat capacity and entropy as a function of temperature for the dipolar spin-ice Hamiltonian with three representative sets of exchange couplings, using both single-spin-flip and loop-update dynamics. The results are plotted in \cref{sfig: dipolar thermo}. 
All three models show a very similar pattern: single-spin-flip data show a single Schottky peak at $T\sim D$ with residual entropy consistent with either dipolar-quadrupolar spin ice or the ideal twenty-vertex model \eqref{seq: ideal ice}.
This Schottky peak remains virtually unchanged in loop-update dynamics, but the residual entropy is released in a low-temperature transition.

Curiously, no sharp feature corresponding to this transition appears in \cref{sfig: dipolar thermo}b. This is also the case for all $0<J_2<0.6D, J_2'=0.05D$, while $J_2'=0$ always shows a sharp transition. This suggests that the difference is due to the underlying ordered phase, FM 2-$k$ and AFM 2-$k$, respectively.

\clearpage
\section{Neutron-scattering patterns of dipolar spin ice}

\begin{figure}
    \centering
    \includegraphics{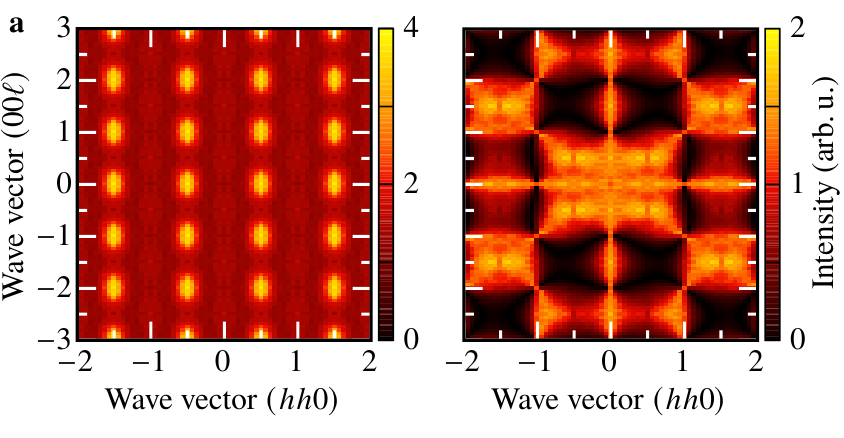}\quad
    \includegraphics{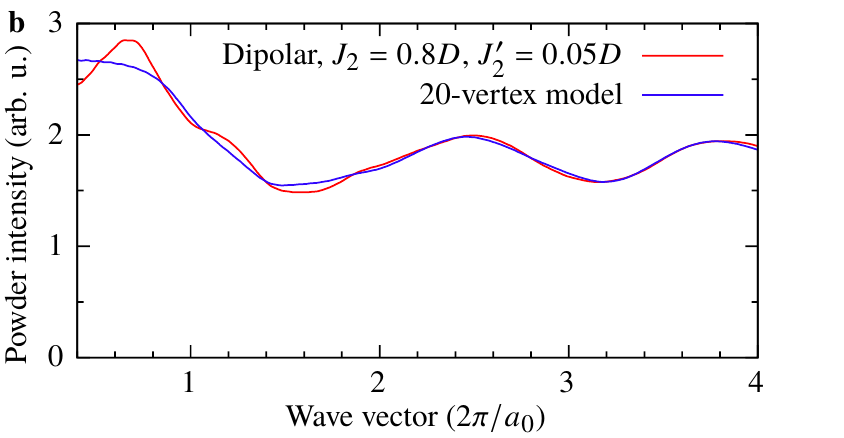}
    \includegraphics{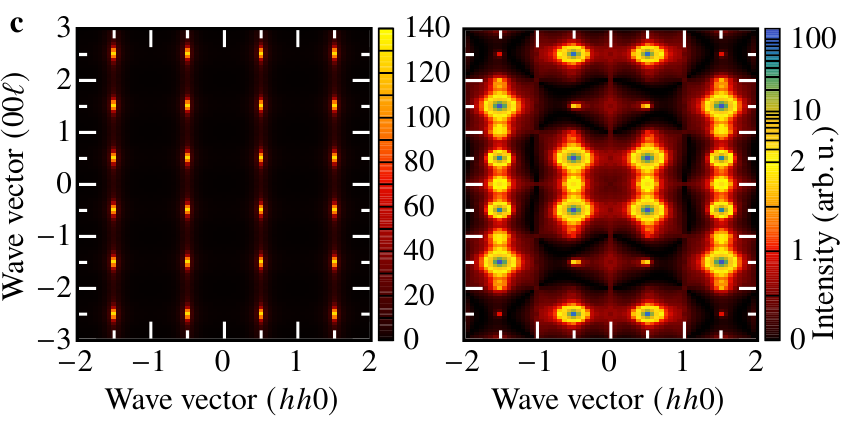}\quad
    \includegraphics{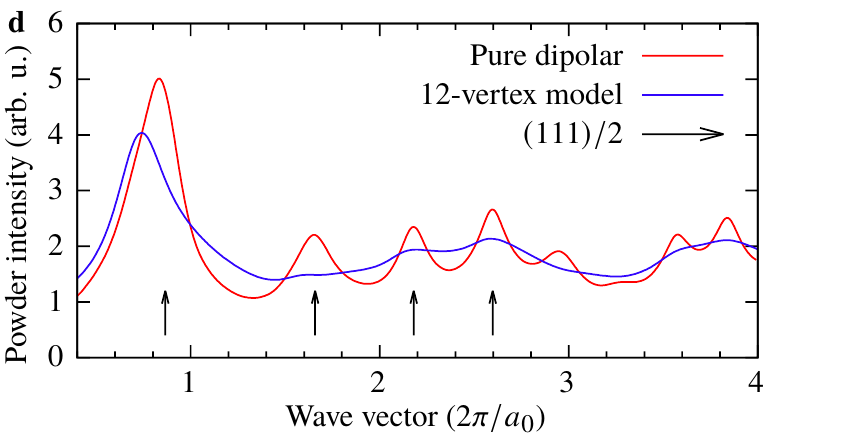}
    \includegraphics{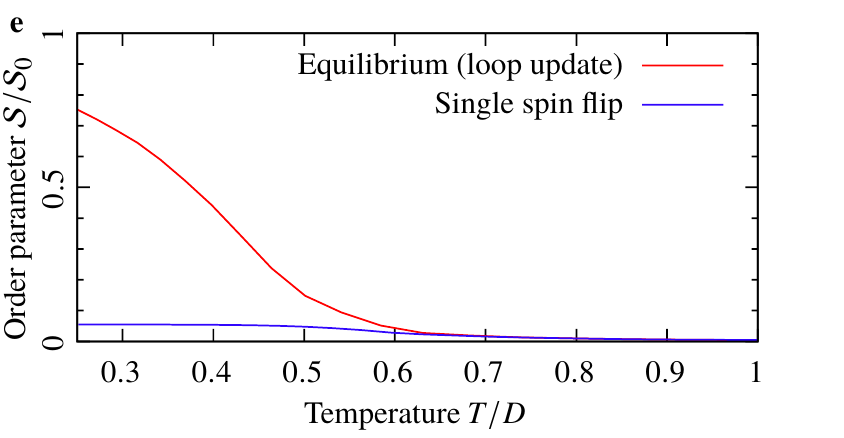}\quad
    \includegraphics{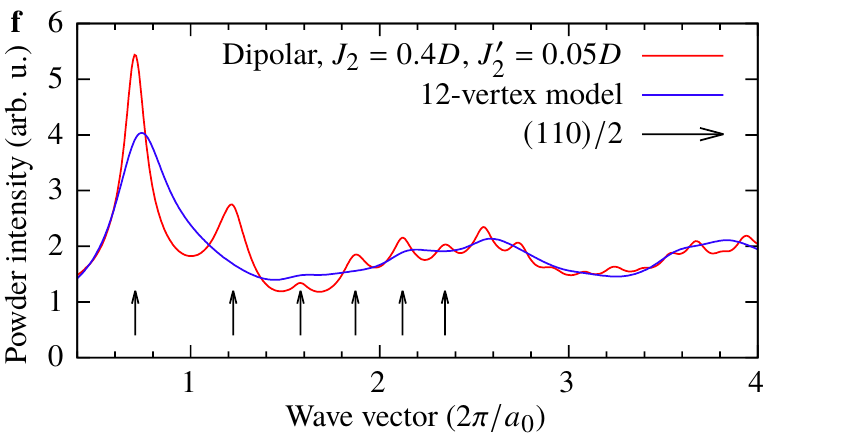}
    \caption{(a,c) Polarised neutron-scattering pattern of the dipolar ice Hamiltonian (a) at $J_2=0.8D$, $J_2'=0.05 D$ and (c) for the pure dipolar Hamiltonian below the glass transition, obtained from single-spin-flip dynamics. Neutrons are polarised along $[1\bar10]$. The spin-flip pattern (right) retains sharp pinch points in both cases. The non-spin-flip pattern (left) shows smooth maxima near $(1/2,1/2,0)$ in case (a) or broadened peaks at $(1/2,1/2,1/2)$ for the pure dipolar model. In the latter case, similar peaks appear in the SF channel as well (note the nonlinear colour scale).
    (e) FM 2-$k$ quadrupolar order parameter $\mathcal{S}^{zz}(0,1/2,1/2)$ as a function of temperature for $J_2=0.4D$, $J_2'=0.05D$ in the dipolar ice Hamiltonian. Loop-update dynamics (red) remains in equilibrium and undergoes an ordering transition at $T\approx0.6D$. Single-spin-flip dynamics falls out of equilibrium at around the same temperature, resulting in a frozen state with small ordered moment and relatively short correlation length, consistent with Fig.~4 of the main text.
    (b,d,f) Powder neutron-scattering pattern of the twenty-vertex model~\eqref{seq: ideal ice}, the twelve-vertex model of dipolar-quadrupolar octahedra, and their dipolar approximations in single-spin-flip dynamics (parameters in legend). 
    Panels (e,f) correspond to the single-crystal patterns shown in Fig. 4 of the main text.
    }
    \label{sfig: NS ice}
\end{figure}

In addition to Fig.~4 of the main text, we obtained polarised neutron-scattering patterns of the dipolar spin ice model both in the high-entropy region $J_2\approx J_2^*$ (\cref{sfig: NS ice}a) and for the pure dipolar Hamiltonian with $J_2=J_2'=0$ (\cref{sfig: NS ice}c). 
The spin-flip pattern in the first case shows prominent pinch points and is altogether very similar to that of the twenty-vertex model~\eqref{seq: ideal ice}. 
The non-spin-flip channel is no longer perfectly flat: the maxima at $(1/2,1/2,0)$ in \cref{sfig: NS ice}a are related to the equilibrium FM 2-$k$ order, so they are liable to change drastically upon tuning near-neighbour exchange constants, similar to pyrochlore spin ice~\cite{Chung2022FlatBandNS}.

Indeed, both the SF and NSF patterns shown in \cref{sfig: NS ice}c show prominent but broad peaks, similar to those in Fig.~4 or the main text, at $(1/2,1/2,1/2)$. The former do not interfere with the pinch points at the $\Gamma$ points, which remain sharp and retain most of their intensity; nevertheless, the pattern looks qualitatively different from that in Fig.~4. This is because the two models fall into different equilibrium phases (cf. Fig.~3b of the main text); the observed broad peaks reflect magnetic reflections due to these orders (cf. \cref{sfig: neutron ordered}c,d).

Powder averages of the neutron-scattering intensity are smooth for both the twenty-vertex and the dipolar-octupolar twelve-vertex model (blue curves in \cref{sfig: NS ice}b,d,f); both are dominated by a slow wavelike modulation on top of a large constant background.
In particular, there are no sharp features associated with pinch points at reciprocal lattice vectors: 
such singularities contribute a $q^2\log|q|$ term to the powder average, too subtle to be noticeable in the graph.
The main difference between the two patterns is a large maximum at $ka/2\pi\approx0.8$, only present in the twelve-vertex case, which is a consequence of the intensity concentration at $(1/2,1/2,k)$ seen in Fig.~1c of the main text.

Dipolar spin ice in the highest-entropy regime (red in \cref{sfig: NS ice}b) follows the twenty-vertex pattern very closely; the only noticeable deviation is a broad maximum at $ka/2\pi\approx 0.7$, consistent with the broad maxima in \cref{sfig: NS ice}a.
By contrast, the powder patterns in the fragmented ice regime (red in \cref{sfig: NS ice}d,f) show prominent but rather broad peaks at either $(1/2,1/2,1/2)$ or $(1/2,1/2,0)$ points (black arrows), consistent with the peaks in the NSF channel of either \cref{sfig: NS ice}c or Fig.~4 in the main text.
The presence and broadness of these peaks indicates that the quadrupolar components of octahedra start to order in the FM 2-$k$ pattern (cf.~\cref{ssec: transition}), but fail to do so completely.
Indeed, the corresponding order parameter (\cref{sfig: NS ice}e) indicates a transition at $T\approx 0.6D$ in equilibrium.
However, the single-spin-flip dynamics starts to become noticeably glassy at the same temperature, resulting in a finite but small frozen order parameter and a short correlation length. This leaves the dipolar components largely unaffected, so Coulomb-phase pinch points remain observable in Fig.~4 of the main text.

\clearpage
\section{Ordered phases}

\subsection{Symmetry analysis of the ordered phases}

\begin{table}
    \centering
    \setlength{\tabcolsep}{1.25ex}
    \begin{tabular}{ccccc}\hline\hline
        phase & magnetic space group & wave vector & irrep & order parameter direction \\\hline
        3-$k$ & $Im\overline{3}m'$ & $(1/2,1/2,0),(1/2,0,1/2),(0,1/2,1/2)$ & $mM_3^+$  & $(a,a,a)$ \\[2pt]\hline
        FM & $R\overline{3}m'$ & $(0,0,0)$ & $m\Gamma_{4}^{+}$ & $(a,a,a)$ \\[2pt]\hline
        \multirow{2}{*}{FM 2-$k$} & \multirow{2}{*}{$P4/nb'm'$} & $(0,0,0)$ & $m\Gamma_4^+$ & $(0,0,a)$\\
        && $(1/2,1/2,0)$& $mM_2^-$ & $(b,0,0)$ \\[1pt]\hline
        \multirow{2}{*}{AFM 2-$k$} & \multirow{2}{*}{$P_{C}4_{2}/mnm$} & $(1/2,1/2,0)$ & $mM_3^+$ & $(a,0,0)$\\
        && $(1/2,1/2,1/2)$ & $mR_3^+$ & $(b,0)$ \\[1pt]\hline
    \end{tabular}
    \caption{Magnetic space groups in the Belov--Neronova--Smirnova notation~\cite{OG1965}, propagation vectors, irreducible representations in the Miller--Love notation~\cite{Miller1967}, and order parameter directions (as returned by \textsc{isodistort}~\cite{Campbell2006}) for the ordered phases described in the main text. For the 2-$k$ structures, the first and second rows give symmetry properties of the dipolar and quadrupolar components, respectively.}
    \label{stab: irreps}
\end{table}

The magnetic symmetry of the four ordered phases found by the Monte Carlo simulations has been determined by group theory analysis.
Calculations were performed using the \textsc{isodistort} and \textsc{isotropy} programs \cite{Campbell2006,Hatch2003}. 
The cubic antiperovskite parent structure belongs to the space group $Pm\overline{3}m1'$; magnetic atoms sit in the $3d$ Wyckoff position $(1/2,0,0)$; the $A$ and $B$ sites correspond to the $1b$ $(1/2,1/2,1/2)$ and $1a$ $(0,0,0)$ positions, respectively.
The symmetry analysis results are summarised in \cref{stab: irreps} and the supplementary mcif files, which provide symmetry elements and atomic positions.
Moreover, group theory calculations indicate that both the Landau and Lifshitz conditions are met, that is, by symmetry, all transitions are allowed to be continuous.

\clearpage
\subsection{Neutron-scattering patterns}

\begin{figure}
    \centering
    \includegraphics{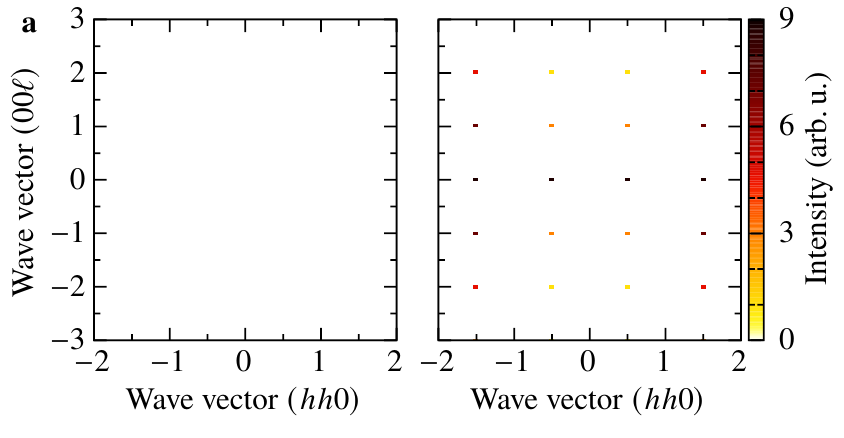}
    \includegraphics{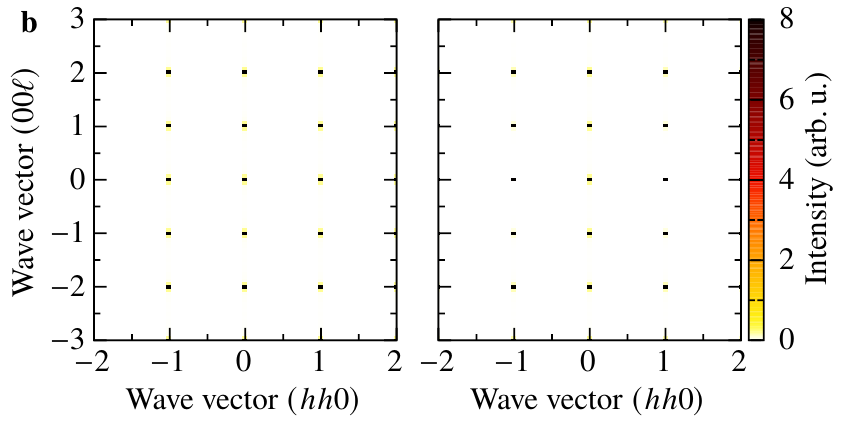}
    \includegraphics{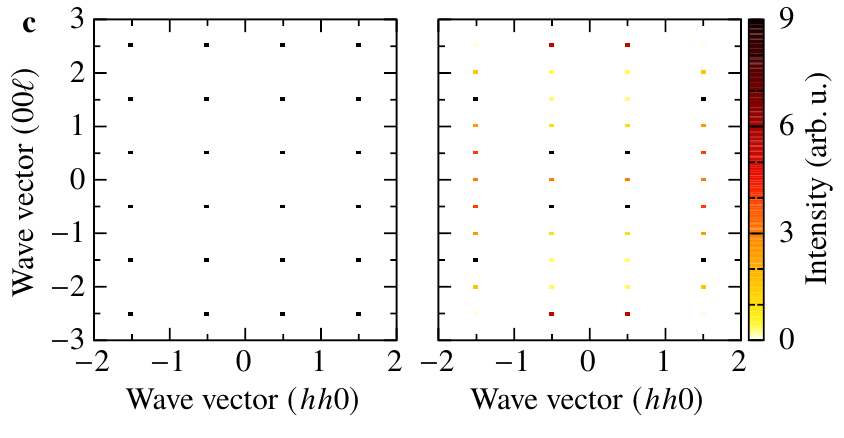}
    \includegraphics{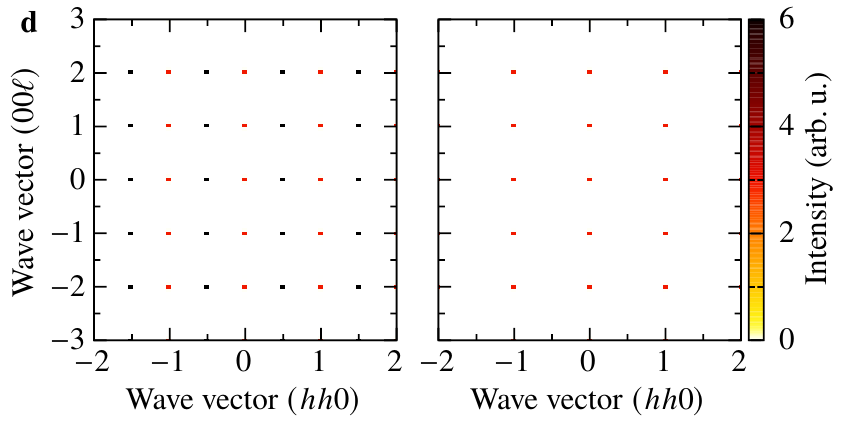}
    \includegraphics{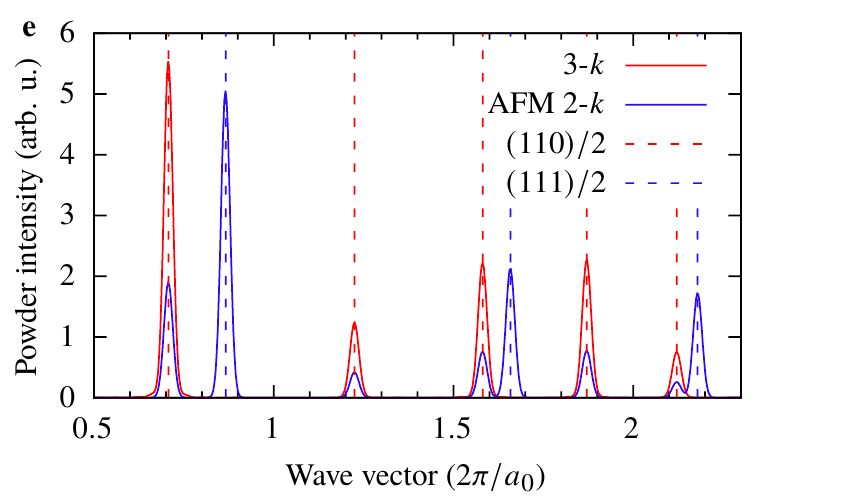}
    \includegraphics{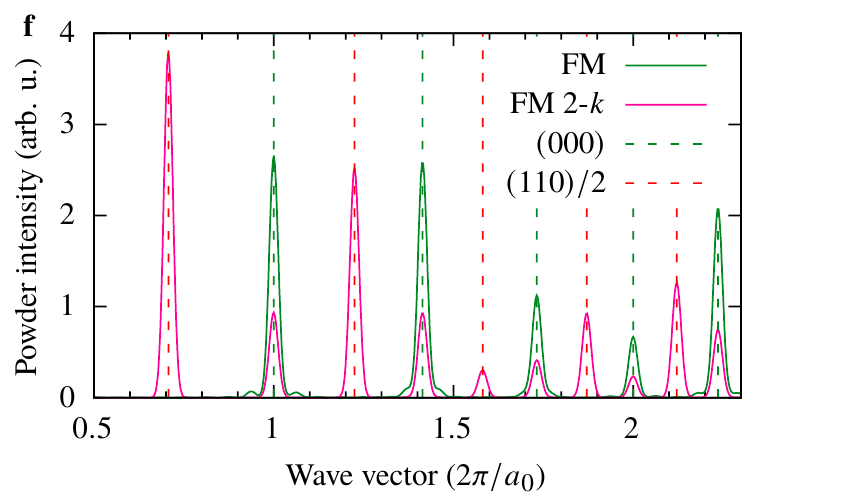}
    \caption{(a-d) Non-spin-flip (left panels) and spin-flip (right panels) neutron-scattering patterns of the four ordered phases in the $(hhk)$ plane (neutrons polarised along $[1\bar10]$). (a) 3-$k$ order ($J_2=D$, $J_2'=-0.2D$); (b) canted ferromagnetic order ($J_2=D$, $J_2'=0.3D$); (c) AFM 2-$k$ order ($J_2=0$, $J_2'=-0.2D$); (d) FM 2-$k$ order ($J_2=0$, $J_2'=0.3D$). The ``arbitrary units'' are equivalent on all graphs.
    (e-f) Powder neutron-scattering patterns of the four ordered phases. All broadening of the peaks is due to the finite simulation box. }
    \label{sfig: neutron ordered}
\end{figure}

Polarised single-crystal and unpolarised powder neutron-scattering patterns for the four ordered phases discussed in the main text were obtained from loop-update Monte Carlo simulations and plotted in Fig.~\ref{sfig: neutron ordered}. All structure factors except the one for the ferromagnetic order are perfectly concentrated at the high-symmetry ordering wave vectors; in the ferromagnetic case, some residual intensity survives along the $(00k)$ line, indicating a finite correlation length along the ferromagnetically aligned chains. We believe this is an artefact of the loop-update Monte Carlo algorithm: in perfect ferromagnetic alignment, only system-spanning flippable loops survive, which makes it hard to reach this state in Monte Carlo dynamics.

\clearpage
\subsection{Nature of the ordering transitions}
\label{ssec: transition}

\begin{figure}
    \centering
    \includegraphics{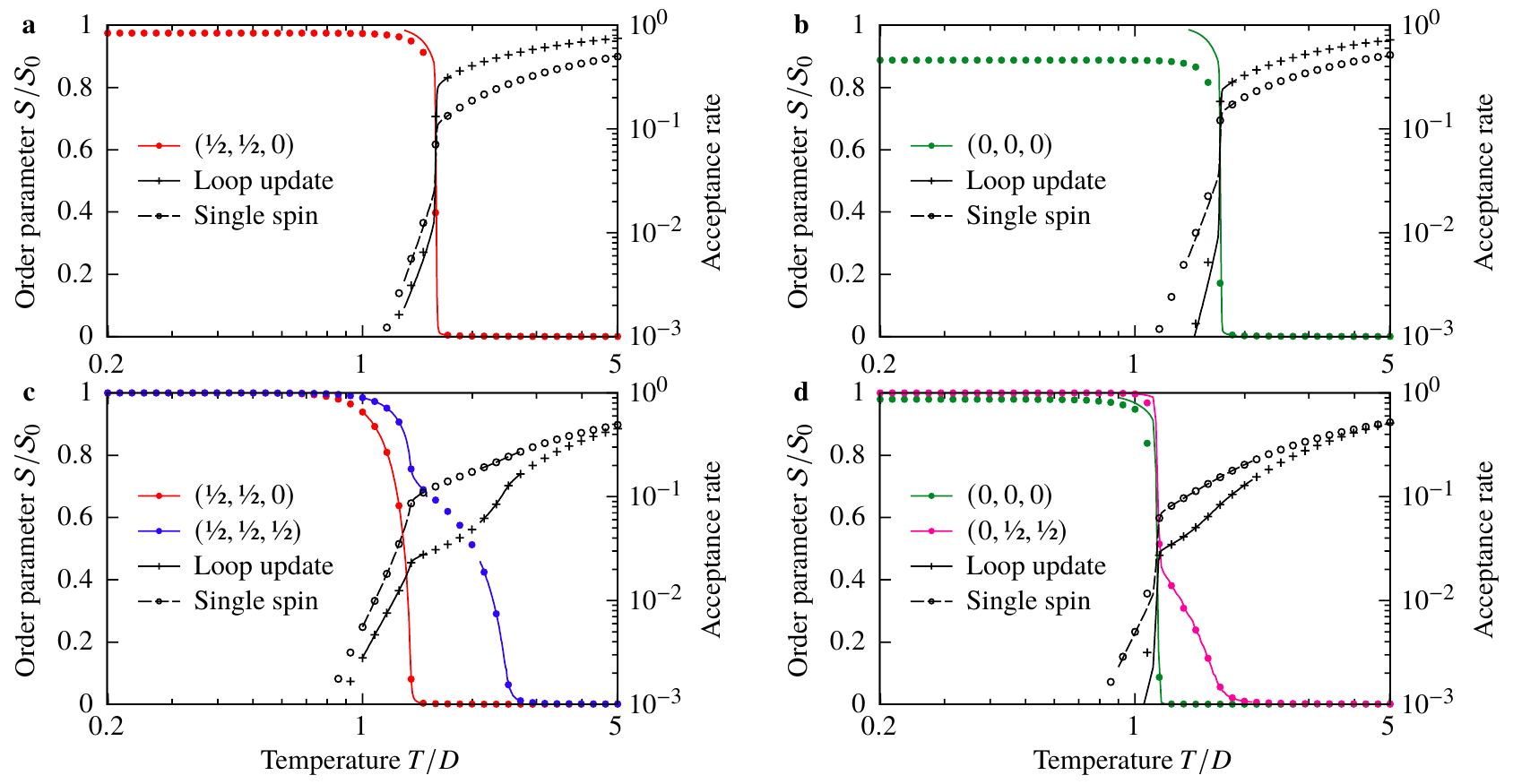}
    \caption{Order parameters (colour) and acceptance rates (black) as a function of temperature at four points deep in the ordered phases: (a) 3-$k$ order ($J_2=D$, $J_2'=-0.2D$); (b) ferromagnetic order ($J_2=D$, $J_2'=0.3D$); (c) AFM 2-$k$ order ($J_2=0$, $J_2'=-0.2D$); (d) FM 2-$k$ order ($J_2=0$, $J_2'=0.3D$). Points show data from the large-span temperature scans in~\cref{stab: Monte Carlo}, lines those from the higher-resolution scans in~\cref{stab: MC transition}. All curves are normalised to the correlator expected in the perfectly ordered structures shown in Fig.~5 of the main text; the colours of the curves match the colouring of magnetic moments in the same figure. }
    \label{sfig: order parameters}
\end{figure}

\def\cS{\mathcal{S}}
The natural order parameter for magnetic ordering transitions is the spin structure factor $\cS^{\alpha\alpha}(\vec k) = \langle S^\alpha(\vec k) S^\alpha(-\vec k)\rangle$.
In particular, the ordered phases seen in our model are captured by the following four order parameters:
\begin{align}
    \text{3-$k$ dipolar:}&\ \cS^{zz}(1/2,1/2,0); &
    \text{FM dipolar:} &\ \cS^{zz}(0,0,0);\nonumber\\*
    \text{AFM 2-$k$ quadrupolar:}&\ \cS^{zz}(1/2,1/2,1/2);&
    \text{FM 2-$k$ quadrupolar:}&\ \cS^{zz}(0,1/2,1/2),
\end{align}
and equivalent ones related to these by cubic rotations.
Indeed, as 2-$k$ ordered phases break cubic rotation symmetry, symmetry-averaging the correlators is necessary to obtain good order parameter statistics (cf.~\cref{ssec: Monte Carlo}).
In addition, we calculated the acceptance rate of both single-spin-flip and short-loop Monte Carlo updates, as these are known to drop suddenly at a first-order transition~\cite{Melko2004MonteModel}.
This data is shown in \cref{sfig: order parameters}.
In the canted ferromagnetic and 3-$k$ phases we see a single, strongly first-order transition, evidenced both by a jump of the relevant order parameter and a sudden drop (by about an order of magnitude) of both acceptance rates. 
On the other hand, the 2-$k$ orders develop in two stages: first, the quadrupolar components order, followed by the dipolar ones.

\begin{table}
    \centering
    \setlength{\tabcolsep}{1.25ex}
    \begin{tabular}{cccccccc} \hline\hline
        $J_2/D$ & $J_2'/D$\strut & $T_\mathrm{max}/D$ & $T_\mathrm{min}/D$ & $\Delta T/D$ & \#burn-in & \#samples & \#rounds \\[2pt] \hline
        1.0 & $-0.2$ & 1.75 & 1.30 & \multirow{4.8}{*}{0.01} & \multirow{6}{*}{128} & \multirow{6}{*}{512} & \multirow{6}{*}{48} \\\cline{1-4}
        1.0 & $+0.3$ & 1.90 & 1.40 &\\\cline{1-4}
        \multirow{2}{*}{0.0} & \multirow{2}{*}{$-0.2$} & 2.70 & 2.10 \\\cline{3-4}
        && 1.50 & 1.00 \\\cline{1-5}
        0.0 & $+0.3$ & 2.10 & 0.90 & 0.02 \\\hline
    \end{tabular}
    \caption{Details of loop-update Monte Carlo simulations in the vicinity of ordering transitions.}
    \label{stab: MC transition}
\end{table}

To better establish the order of these transitions, we performed additional Monte Carlo simulations with finer temperature resolution near the ordering temperatures. 
Details of these calculations are summarised in \cref{stab: MC transition} and the results are shown in \cref{sfig: order parameters} in solid lines.
These results clearly show that the higher-temperature transition of quadrupolar components is continuous (neither order parameters nor acceptance rates change abruptly), while all dipolar transitions are again first-order, with the possible exception of the low-temperature transition in \cref{sfig: order parameters}c.
The latter is also emphasised by the mismatch between the data with finer and coarser $T$-resolution, which hints at slow nucleation dynamics.

\bibliography{paper,references}